\documentclass[prletter,showpacs,twocolumn,typeset,superscriptaddress]{revtex4}  
\usepackage{amsfonts}
\usepackage{amsmath}
\usepackage{amssymb}
\usepackage{graphicx}
\usepackage{epsfig}
\usepackage{citesort}

\usepackage{dcolumn}
\usepackage{bm}

\begin{document}

\preprint{APS}

\title{Narrow entanglement beats}
\author{Luis Roa}
\affiliation{Center for Quantum Optics and Quantum Information, Departamento de F\'{\i}sica,
Universidad de Concepci\'{o}n, Casilla 160-C, Concepci\'{o}n,
Chile.}
\author{R. Pozo-Gonz\'{a}lez}
\affiliation{Departamento de F\'{\i}sica, Tecnol\'ogico de Monterrey, Monterrey 64849, Mexico.}
\affiliation{Center for Quantum Optics and Quantum Information, Departamento de F\'{\i}sica,
Universidad de Concepci\'{o}n, Casilla 160-C, Concepci\'{o}n,
Chile.}
\author{Marius Schaefer}
\affiliation{Center for Quantum Optics and Quantum Information, Departamento de F\'{\i}sica,
Universidad de Concepci\'{o}n, Casilla 160-C, Concepci\'{o}n,
Chile.}
\affiliation{Eidgen\"{o}ssisches Institut f\"{u}r Schnee und Lawinenforschung, Fl\"{u}elastrasse 11, CH-7260, Davos Dorf, Switzerland}
\author{P. Utreras-SM}
\affiliation{Center for Quantum Optics and Quantum Information, Departamento de F\'{\i}sica,
Universidad de Concepci\'{o}n, Casilla 160-C, Concepci\'{o}n,
Chile.}

\date{\today}

\begin{abstract}
We study how the entanglement between two atoms can be created or modified even when they do not interact
but when each of them interacts dispersively, i.e., weak and far from the resonance with a single mode of the field.
Considering that regime we apply a method which makes use of a small nonlinear deformation of the usual $SU(2)$ algebra in order to obtain the effective Hamiltonian describing correctly the dynamics for any initial states.
In particular we study two cases: In the first one we consider each atom initially in a pure state and in the second case we assume that they start in a Werner state.
We find that both atoms can reach, periodically, maximum entanglement if each of them starts in any eigenstate of $\sigma_x$, independent of the initial Fock state of the mode.
Thus we find that a dispersive vacuum can generate entanglement between two two-level atoms.
In the second case and when the field mode is initially in a coherent or thermal state, we find that in the high energy limit, in general,
there is no entanglement between the two atoms however at well defined moments the initial entanglement is as suddenly recovered as removed.
This time behavior looks like narrow beats separated by the so called \textit{entanglement dead valleys}.
\end{abstract}

\maketitle

\section{Introduction}
In 1935 E. Schr\"{o}dinger introduced into the quantum world the entanglement concept by means of his communication
addressing the gedanken experiment known as Schr\"{o}dinger's cat \cite{Schrodinger}.
In the same year A. Einstein, B. Podolsky, and N. Rosen argued the incompleteness of quantum mechanics by describing the reality they sensed \cite{Einstein}.
Later, in 1964, J. S. Bell reported that a special no local operator must satisfy in average an inequality which can be violated only by some non separable states \cite{Bell}.
Today the non locality effect or entanglement is considered as a resort for manipulation of quantum information which does not have a classical counterpart \cite{Bennett,Zoller}.
Thus, during the last two decades a major research effort has been conducted in the emerging field of quantum information theory \cite{Nielsen} based on the renewed non locality effect.
With this motivation, there has been a lot of interest in understanding and quantifying entanglement of pure and mixed states \cite{Caves,Vedral,DiVincenzo,Wootters}.
The entanglement of two systems can arise through direct interaction between them as well as through the coupling of the systems with a common quantum bus in the form of an auxiliary system or environment \cite{Tessier,Steinbach,Sainz,Bose,Eberly,Davidovich}.

In this work we investigate whether and how the entanglement between two atoms can be generated or modified when they interact dispersively with the same single mode of the electromagnetic field.
We use the method of the small nonlinear deformation of the usual $SU(2)$ algebra \cite{Klimov,Klimov2,Gottfried},
in order to obtain an effective Hamiltonian describing the dynamics of the system.

\section{The Hamiltonian Model}

We consider two noninteracting two-level atoms, labelled by sub or supraindexes $a$ and $b$, each one coupled dispersively with a single mode characterized by the frequency $\omega$.
The unitary dynamics in the whole tensorial product Hilbert space, $\textsc{H}=\textsc{H}_{a}\otimes \textsc{H}_{b}\otimes\textsc{H}_{\text{mode}}$, is driven by the Hamiltonian ($\hbar =1$) in the rotating wave approximation,
\begin{eqnarray}
\hat{H}&=&\frac{1}{2}\omega _a\sigma_z^{(a)}+\frac{1}{2}\omega_b\sigma_z^{(b)}+\omega b^{\dagger}b \nonumber \\
&&+g_a(\sigma_+^{(a)}b+\sigma_-^{(a)}b^{\dagger})
+g_b(\sigma_+^{(b)}b+\sigma_-^{(b)}b^{\dagger}),              \label{H}
\end{eqnarray}
where $b$ and $b^{\dagger }$ are the annihilation and creation single mode operators,
$\sigma _{+}^{(j)}=|1\rangle _{j{}j}\langle 0|$,
$\sigma_{-}^{(j)}=|0\rangle _{j{}j}\langle 1|$, and $\sigma _{z}^{(j)}$ is the $z-$
component of the effective angular spin-half operator whose eigenstates are $\{|0\rangle _{j},|1\rangle _{j}\}$, for $j=a,b$.

Taking into account that the excitation number operator $\hat{N}=(\sigma _{z}^{(a)}+\sigma _{z}^{(b)})/2+b^{\dagger}b$ is a constant of motion, $[\hat{H},\hat{N}]=0$, the (\ref{H}) Hamiltonian can be written by $\hat{H}=\omega\hat{N}+\hat{H}_{int}$ where
\begin{eqnarray}
\hat{H}_{int}&=&\frac{\Delta_a}{2}\sigma_z^{(a)}+\frac{\Delta_b}{2}\sigma_z^{(b)}    \nonumber \\
&&+g_a(\sigma_+^{(a)}b+\sigma_-^{(a)}b^{\dagger})+g_b(\sigma_+^{(b)}b+\sigma_-^{(b)}b^{\dagger}), \label{Hint}
\end{eqnarray}
with $\Delta _{a}=\omega _{a}-\omega$ and $\Delta _{b}=\omega _{b}-\omega$.

We have assumed dispersive interactions between each atom and the common single mode.
In other words, those couplings are weak and far from the resonance, so we can define the small parameters:
\begin{equation}
\epsilon_j\equiv\frac{g_j}{\Delta_j}\ll\frac{1}{\sqrt{\langle n\rangle_T}}\ll 1,\hspace{0.5in}j=a,b,
\end{equation}
where $\langle n\rangle_T$ is the average photon number.
Making use of the small rotation method \cite{Klimov,Klimov2} to obtain the effective Hamiltonian which approximately describes the interaction process, we can eliminate the two terms which do not represent the resonance interaction but represent rapid oscillations in the rotating frame.
That can be achieved by applying to the Hamilnonian (\ref{Hint}) a small unitary transformation:
\begin{equation}
\hat{R}=e^{\epsilon_a(\sigma_+^{(a)}b-\sigma_-^{(a)}b^{\dagger})+\epsilon_b(\sigma_+^{(b)}b-\sigma_-^{(b)}b^{\dagger})}. \nonumber
\end{equation}
Considering terms up to first order in $\epsilon_a$ and $\epsilon_b$ of the Cambell-Baker-Hausdorf expansion, $\hat{R}H_{int}\hat{R}^\dagger$, we obtain the following effective Hamiltonian:
\begin{eqnarray}
\hat{H}_{eff}&=&
\frac{\Delta_a}{2}\sigma_z^{(a)}+\frac{\Delta_b}{2}\sigma_z^{(b)}   \nonumber \\
&&+(b^{\dagger}b+\frac{1}{2})\left(\frac{g_a^2}{\Delta_a}\sigma_z^{(a)}+\frac{g_b^2}{\Delta_b}\sigma_z^{(b)}\right)\nonumber \\
&&+\frac{g_ag_b}{2}(\frac{1}{\Delta_a}+\frac{1}{\Delta_b})\left(\sigma_+^{(a)}\sigma_-^{(b)}+\sigma_-^{(a)}\sigma_+^{(b)}\right). \label{Heff}
\end{eqnarray}
The first two terms in the above effective Hamiltonian represents the free evolution of the non-resonant atoms with renormalized transition frequencies.
The third term is the so-called, dynamical Stark shifts, which describes an additional intensity dependent detuning of non-resonant atoms from the mode frequency.
The last term represents an effective dipole-dipole interaction between the non-resonant atoms which appears as a consequence of a collective nature of the
interaction of non-resonant atoms with the quantized mode.
We point out that this kind of effective interaction could not appear in the classical field and that the contribution of this term strongly depends on the
internal resonance conditions of the non-resonant atoms.

The effective Hamiltonian (\ref{Heff}) can be diagonalized without difficulty but, for the sake of simplicity, we suppose that both atoms are identical in a way such that $g=g_a=g_b$ and $\Delta=\Delta_a=\Delta_b$.
Under those conditions the unitary evolution operator is given by
\begin{eqnarray}
\hat{U}&=&e^{-i[\Delta +g^2(2b^{\dagger}b+1)/\Delta]t}|1\rangle_a|1\rangle_{b\ a}\langle 1|_b\langle1|    \nonumber \\
&&+e^{-ig^2t/\Delta}|\psi^+\rangle\langle\psi^+|+e^{ig^2t/\Delta }|\psi^-\rangle\langle\psi^-|    \nonumber  \\
&&+e^{i[\Delta+g^2(2b^{\dagger}b+1)/\Delta]t}|0\rangle_a|0\rangle_{b\ a}\langle 0|_b\langle 0|,
\label{U}
\end{eqnarray}
where $|\psi^{\pm}\rangle=(|0\rangle_a|1\rangle_b\pm|1\rangle_a|0\rangle_b)/\sqrt{2}$ are two Bell states.

From Eqs. (\ref{Heff}) and (\ref{U}) one realizes that the dressed states $|1\rangle_a|1\rangle_b|n\rangle$, $|0\rangle_a|0\rangle_b|n\rangle$, and $|\psi^{\pm}\rangle|n\rangle$ ($n=0,1,2,\dots$) are stationary. Thus, by their form, independent of the initial mode state, the $a-b$ bipartite system does not evolve when starting at one of the states $|1\rangle_a|1\rangle_b$,
$|0\rangle_a|0\rangle_b$, or $|\psi^{\pm}\rangle$, and hence preserves the initial entanglement.

\section{Bipartite Entanglement}

The entanglement between two systems in a whole pure state is given by the entropy of any subsystem.
It can also be evaluated by the concurrence $C(|\psi\rangle)\equiv|\langle\psi |\sigma_y\otimes\sigma_y|\psi^*\rangle|$, where the asterisk denotes complex conjugation of the probability amplitudes in the $\sigma_z\otimes\sigma_z$-representation, i.e., in the base: $\{|0\rangle|0\rangle,|0\rangle|1\rangle,|1\rangle|0\rangle,|1\rangle|1\rangle\}$ \cite{Caves,Wootters}.
The generalization of the concurrence to a mixed state $\rho$ of two atoms is defined as the infimum of the average concurrence over all possible pure state ensemble decompositions of $\rho$, defined as convex combinations of pure states $s_i=\{p_i,|\psi_i\rangle\}$ decomposition, such that $\rho=\sum_ip_i|\psi_i\rangle\langle\psi_i|$.
In this way, $C(\rho)=\inf_{s_i}\sum_ip_iC(|\psi_i\rangle)$.
Williams Wootters succeeded in deriving an analytic solution to this difficult minimization procedure in terms of the eigenvalues $\lambda_i$'s of the non-Hermitian
operator $\rho\sigma_y\otimes\sigma_y\rho^*\sigma_y\otimes\sigma_y$, where the asterisk again denotes complex conjugate of the elements of $\rho$ in the $\sigma_z\otimes\sigma_z$-representation.
The closed-form solution for the concurrence of a mixed state of two atoms is given by $C(\rho)=\max\{0,\sqrt{\lambda_1}-\sqrt{\lambda_2}-\sqrt{\lambda_3}-\sqrt{\lambda_4}\}$,
where the $\lambda_i$'s are in the decreasing order\cite{Wootters}.

First we study the entanglement of formation between the two atoms when each of them is initially in a pure state.
After that we consider the two atoms in a mixed states.
In both cases we study  how the entanglement evolves as a function of time and energy mode, under the weak and dispersive interaction (\ref{Heff}).
We consider the initial mode state to be a Fock state $|n\rangle$, a coherent state $|\alpha\rangle$, and a thermal state $\rho_T=\sum_n \langle n\rangle_T^n/(1-\langle n\rangle_T)^{n+1}|n\rangle\langle n|$,
where $\langle n\rangle_T=1/(e^{\omega/k_BT}-1)$ is the average photon number of the mode, $T$ the absolute temperature and $k_B$ the Boltzmann constant ($\hbar=1$).

\subsection{Initial pure state}

Here we consider each atom to be initially in a pure state, i.e., qubit $a$ being in $|\psi\rangle_a$ and qubit $b$ in $|\varphi\rangle_b$.
When the single mode is initially in a Fock state $|n\rangle$, the two atoms-system does not tangle with it at all time, and they evolve to the following pure state:
\begin{eqnarray}
|\phi\rangle &=&\langle 0|\psi\rangle\langle 0|\varphi\rangle e^{i[\Delta+\frac{g^2}{\Delta}(2n+1)]t}|0\rangle_a|0\rangle_b
+|L\rangle            \nonumber \\
&&+\langle 1|\psi\rangle\langle 1|\varphi\rangle e^{-i[\Delta+\frac{g^2}{\Delta}(2n+1)]t}|1\rangle_a|1\rangle_b,
\nonumber
\end{eqnarray}
where $|L\rangle =L_0|0\rangle_a|1\rangle_b+L_1|1\rangle_a|0\rangle_b$ is an unnormalized state in the subspace spanned by $\{|0\rangle_a|1\rangle_b,|1\rangle_a|0\rangle_b\}$.
Here we have defined the functions:
\begin{eqnarray}
L_0&=&\langle 0|\psi\rangle\langle 1|\varphi\rangle\cos\frac{g^{2}t}{\Delta}-i\langle 1|\psi\rangle\langle 0|\varphi\rangle\sin\frac{g^{2}t}{\Delta},
\nonumber   \\
L_1&=&\langle 1|\psi\rangle\langle 0|\varphi\rangle\cos\frac{g^{2}t}{\Delta}-i\langle 0|\psi\rangle\langle 1|\varphi\rangle\sin\frac{g^{2}t}{\Delta}.
\nonumber
\end{eqnarray}
Thus, the concurrence of the above pure state is read as follows
\begin{equation}
C(|\phi \rangle )=2\left\vert L_{0}L_{1}-\langle 0|\psi \rangle \langle 0|\varphi \rangle
\langle 1|\psi \rangle \langle 1|\varphi \rangle \right\vert.
\label{c1}
\end{equation}
From Eq. (\ref{c1}), we can see that the concurrence: i) does not rely on the $n$ photon number,
ii) reaches the maximum value $1$ at $t=\pi\Delta/(2g^2)$ and hence periodically, under the condition that both atoms start in any eigenstate of $\sigma_x$ or even in a general state of type $|\theta\rangle=(|0\rangle+e^{i\theta}|1\rangle)/\sqrt{2}$ being $\theta$ real.
Therefore two atoms starting in any $|\theta\rangle$ state reach maximum entanglement when they interact dispersively even with a common vacuum.
In this form we can say that the vacuum can generate entanglement between two two-level atoms even when they are far from the resonance.

When the single mode is initially in a coherent state $|\alpha\rangle$, the reduced density operator of the two atoms-system becomes,
\begin{widetext}
\begin{eqnarray}
\rho&=&|\langle 1|\psi\rangle\langle 1|\varphi\rangle|^2|1\rangle_a|1\rangle_{ba}\langle 1|_b\langle 1|+|L\rangle\langle L|
+|\langle 0|\psi\rangle\langle 0|\varphi\rangle|^2|0\rangle_a|0\rangle_{ba}\langle 0|_b\langle 0|     \nonumber \\
&&+\langle 0|\psi\rangle\langle 0|\varphi\rangle e^{i(\Delta+g^{2}/\Delta)t}e^{-|\alpha|^{2}(1-e^{2ig^{2}t/\Delta})}|0\rangle_a|0\rangle_b\langle L| \nonumber \\
&&+\langle 0|\psi\rangle\langle\psi|1\rangle\langle 0|\varphi\rangle\langle\varphi|1\rangle e^{2i(\Delta+g^{2}/\Delta)t}e^{-|\alpha|^{2}(1-e^{4ig^{2}t/\Delta})}
|0\rangle_a|0\rangle_{ba}\langle1|_b\langle 1| \nonumber \\
&&+\langle\psi|0\rangle\langle\varphi|0\rangle e^{-i(\Delta+g^{2}/\Delta)t}e^{-|\alpha|^{2}(1-e^{-2ig^{2}t/\Delta})}|L\rangle_a\langle 0|_b\langle 0| \nonumber \\
&&+\langle\psi|1\rangle\langle\varphi|1\rangle e^{i(\Delta+g^{2}/\Delta )t}e^{-|\alpha|^{2}(1-e^{2ig^{2}t/\Delta})}|L\rangle_a\langle 1|_b\langle 1| \nonumber \\
&&+\langle\psi|0\rangle\langle 1|\psi\rangle\langle\varphi|0\rangle\langle 1|\varphi\rangle e^{-2i(\Delta +g^{2}/\Delta )t}e^{-|\alpha|^{2}(1-e^{-4ig^{2}t/\Delta})}
|1\rangle_a|1\rangle_{ba}\langle 0|_b\langle 0| \nonumber \\
&&+\langle 1|\psi\rangle\langle 1|\varphi\rangle e^{-i(\Delta+g^{2}/\Delta)t}e^{-|\alpha|^{2}(1-e^{-2ig^{2}t/\Delta})}|1\rangle_a|1\rangle_b\langle L|,
\label{rhota}
\end{eqnarray}
\end{widetext}
From this expression, Eq. (\ref{rhota}), we see that only the last six non diagonal terms depend on the $|\alpha|$ and they, in general, vanish for $|\alpha|\gg1$.
However, at times $t=t_k=\pi k\Delta/g^2$ ($k=1,2,\dots$) those terms suddenly reappear and become independent of the intensity $|\alpha|$.
Since at times $t_k$ the state described by Eq. (\ref{rhota}) is equal to the state when $\alpha=0$, that is, with an initial vacuum state, then at those times the concurrence is given by the Eq. (\ref{c1}) evaluated at $t_k$ which is zero. We can also see that when the atom $a$ starts in the $|0\rangle$ state and the atom $b$ begins in the $|1\rangle$ state or vice versa the concurrence of mixed state (\ref{rhota}) is given by $|\sin(2g^2t/\Delta)|$ reaching the maximum value $1$ at $t=\pi\Delta/4g^2$ and repeating it periodically.
On the other hand, in the high intensity regime, $|\alpha|\gg1$, the reduced density operator at any $t\neq t_k$, is given by the first three diagonal terms of the Eq. (\ref{rhota}) and its concurrence is read as follows:
\begin{equation}
C(\rho)=\max\{0,2(|L_0L_1|-|\langle 0|\psi\rangle\langle 0|\varphi\rangle\langle 1|\psi\rangle\langle 1|\varphi\rangle|)\}. \label{c2}
\end{equation}
In this regime and when each atom begins in any $|\theta\rangle$ state, the entanglement between them is always zero.

\begin{figure}[t]
\includegraphics[angle=360,width=0.40\textwidth]{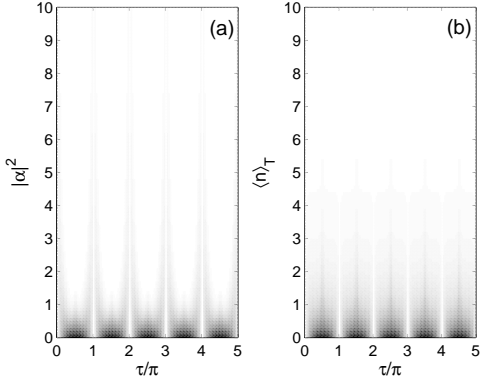}
\caption{Concurrence as a function of the dimensionless time $\tau/\pi$ and of the average photon number of the single mode when the initial mode state is: (a) a coherent state and (b) a thermal state. In both cases each atom starts in the $|\theta=0\rangle$ state. We have considered $g/\Delta=0.01$.
White color means zero entanglement of formation whereas black color stand for their maximum entanglement value 1.}  \label{figure1}
\end{figure}

When the initial mode state is a thermal state at absolute temperature $T$, the reduced density operator of the two atoms, at time $t$ becomes:
\begin{widetext}
\begin{eqnarray}
\rho
&=&|\langle 1|\psi \rangle\langle 1|\varphi \rangle |^{2}|1\rangle_{a}|1\rangle _{ba}\langle 1|_{b}\langle 1|+|L\rangle \langle L|
+|\langle 0|\psi \rangle\langle 0|\varphi \rangle |^{2}|0\rangle _{a}|0\rangle_{ba}\langle 0|_{b}\langle 0|  \nonumber   \\
&&+\frac{e^{i(\Delta +\frac{g^{2}}{\Delta })t}}{1+\langle n\rangle_T(1-e^{i\frac{2g^{2}t}{\Delta }})}(\langle 0|\psi \rangle \langle 0|\varphi \rangle |0\rangle_{a}|0\rangle _{b}\langle L|
+\langle \psi |1\rangle \langle \varphi|1\rangle |L\rangle _{a}\langle 1|_{b}\langle 1|)     \nonumber  \\
&&+\frac{e^{-i(\Delta +\frac{g^{2}}{\Delta })t}}{1+\langle n\rangle_T(1-e^{-i\frac{2g^{2}t}{\Delta }})}(\langle 1|\psi \rangle \langle 1|\varphi \rangle
|1\rangle_{a}|1\rangle _{b}\langle L|+\langle \psi |0\rangle \langle\varphi |0\rangle |L\rangle _{a}\langle 0|_{b}\langle 0|)     \nonumber   \\
&&+\frac{\langle 0|\psi \rangle \langle 0|\varphi \rangle \langle \psi|1\rangle \langle \varphi |1\rangle e^{2i(\Delta +\frac{g^{2}}{\Delta })t}}
{1+\langle n\rangle_T(1-e^{i\frac{4g^{2}t}{\Delta }})}|0\rangle _{a}|0\rangle_{ba}\langle 1|_{b}\langle 1|   \nonumber    \\
&&+\frac{\langle 1|\psi \rangle \langle 1|\varphi \rangle \langle \psi|0\rangle \langle \varphi |0\rangle e^{-2i(\Delta +\frac{g^{2}}{\Delta })t}}
{1+\langle n\rangle_T(1-e^{-i\frac{4g^{2}t}{\Delta }})}|1\rangle _{a}|1\rangle_{ba}\langle 0|_{b}\langle 0|.
\label{rhoTER} 
\end{eqnarray}
\end{widetext}
Once again we find that at high intensity, i.e., at high temperature, $\langle n\rangle_T\gg 1$, the last six non diagonal terms vanish and reappear suddenly at times $t=t_k$.
Like the previous case at those $t_k$ times the concurrence is zero and at any $t\neq t_k$ the concurrence is given by Eq. (\ref{c2}).

Figures (\ref{figure1}) shows a linear black-white degradation of the concurrences of the mixed state given by: (a) the Eq. (\ref{rhota}) and (b) the Eq. (\ref{rhoTER}), as a function of the dimensionless time $\tau=2g^2t/\Delta$ and of the average photon number of the single mode.
White color means zero entanglement of formation whereas black color stand for they maximum entanglement value 1.
In both figures (\ref{figure1}) we considered both qubit starting in the $|\theta=0\rangle$ state.

From the figures (\ref{figure1}) we see that maximal entanglement arises periodically at low energy regime.
This effect is a reminiscence of the entanglement generated by the dispersive vacuum state.
Those maximal entanglement zones are separated by narrow \textit{entantaglement dead valleys} (EDV) \cite{Eberly}.

\subsection{Initial mixed state}

Now we study the case when both atoms are initially in a Werner state \cite{Werner,Miranowicz} type:
\begin{equation}
\rho(0)=\frac{1-\gamma}{4}I+\gamma|X\rangle\langle X|,     \label{W}
\end{equation}
with $I$ being the identity of the two atoms Hilbert space, $|X\rangle$ being one of the four Bell states \cite{Miranowicz}, and $0\leq\gamma\leq 1$ a physical parameter.

One can prove easily that when the single mode starts in a Fock state $|n\rangle$ the concurrence does not change and remains in $(3\gamma-1)/2$ for $\gamma\geq1/3$ and zero otherwise.

However, when the field starts in a coherent state $|\alpha\rangle$ the reduced density operator of the two qubit-system at time $t$ becomes
\begin{eqnarray}
\rho&=&\frac{1-\gamma}{4}I+\frac{\gamma}{2}(|0\rangle_a|0\rangle_{ba}\langle 0|_b\langle 0|+|1\rangle_a|1\rangle_{ba}\langle 1|_b\langle 1|\nonumber \\
&&\pm e^{2i\Omega t}e^{-|\alpha|^2(1-e^{\frac{4ig^{2}t}{\Delta}})}|0\rangle_a|0\rangle_{ba}\langle1|_b\langle 1| \nonumber \\
&&\pm e^{-2i\Omega t}e^{-|\alpha|^2(1-e^{-\frac{4ig^{2}t}{\Delta}})}|1\rangle_a|1\rangle_{ba}\langle0|_b\langle 0|),
\label{rhota2}
\end{eqnarray}
where we have consider $|X\rangle$ to be one of the two Bell states $|\phi^\pm\rangle=(|0\rangle_a|0\rangle_b\pm|1\rangle_a|1\rangle_b)/\sqrt{2}$. $\Omega=\Delta+g^2/\Delta$.
The concurrence of the bipartite mixed state (\ref{rhota2}) is given by
\begin{equation}
C(\rho)=\frac{\left( 1+2e^{-2|\alpha |^{2}\sin ^{2}\frac{2g^{2}t}{\Delta }}\right)\gamma -1}{2}, \label{Ca}
\end{equation}
for $\gamma \geq 1/(1+2e^{-2|\alpha |^{2}\sin ^{2}\frac{2g^{2}t}{\Delta }})$ and is zero otherwise.
Clearly we see that for a high intensity coherent state, in general, there is not entanglement between the two atoms however at each time
$t=t_k/2$ the initial entanglement amount, $\max\{0,(3\gamma-1)/2\}$, is suddenly recovered and is independent of the intensity of the coherent state.
When one consider the other two Bell states $|X\rangle=|\psi^\pm\rangle=(|0\rangle_a|1\rangle_b\pm|1\rangle_a|0\rangle_b)/\sqrt{2}$, the (\ref{W}) density operator does not evolve.

\begin{figure}[t]
\includegraphics[angle=360,width=0.40\textwidth]{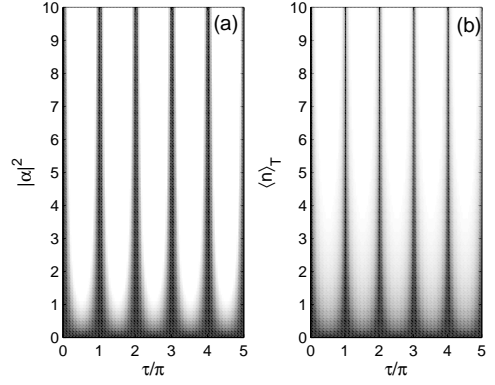}
\caption{Concurrence as a function of the dimensionless time $\tau/\pi$ and of the average photon number of the single mode when the initial mode state is: (a) a coherent state and (b) a thermal state. In both cases the two qubits start in a Werner state.
White color means zero entanglement of formation whereas black color stand for their maximum entanglement value $(3\gamma-1)/2=8/11$ with $\gamma=9/11$.}
\label{figure2}
\end{figure}

On the other hand, when the mode state is initially in a thermodynamic equilibrium at absolute temperature $T$ the concurrence becomes:
\begin{equation}
C(\rho) =\max\{0,\frac{\left(1+\frac{2}{|1+\langle n\rangle_T(1-e^{-4ig^{2}t/\Delta })|}\right) \gamma -1}{2}\}. \label{c3}
\end{equation}
Once again we find the \textit{entanglement-beats} effect at high intensity or equivalently in the high temperature regime, that is, the initial entanglement amount is suddenly recovered just at each times $t=t_k/2$. We will call this effect \textit{E-beats}. 
We can also seen from Eqs. (\ref{c2}) and (\ref{c3}) that when the field mode is initially in the vacuum state, $|\alpha|^2=\langle n\rangle_T=0$, the entanglement does not change.
The expression (\ref{c3}) was calculated considering the state $|X\rangle=|\phi^\pm\rangle$ in Eq. (\ref{W}).
When $|X\rangle=|\psi^\pm\rangle$ the (\ref{W}) density operator does not evolve.

Figures (\ref{figure2}) shows a linear black-white degradation of the concurrences given by: (a) Eq. (\ref{c2}) and (b) Eq. (\ref{c3}), as a function of the dimensionless time $\tau=2g^2t/\Delta$ and of the average photon number of the single mode.
White color means zero entanglement of formation whereas black color stand for they maximum entanglement value $(3\gamma-1)/2=8/11$.
In both figures (\ref{figure2}) we considered $\gamma=9/11$.
From the figures (\ref{figure2}) we see that the \textit{E-beats} effect becomes apparent even for $\langle n\rangle_T=|\alpha|^2 \approx 3$.
These E-beats are separated by EDVs \cite{Eberly}.
  
\section{Conclusions}

In summary we have studied the dynamics of the entanglement between two non interacting two-level atoms weakly coupled and far from the resonance with the same single mode field.
We find that a dispersive vacuum can generate maximum entanglement between them when there is a single photon to share. 
We emphasize that in the dispersive regime, the atomic energy is not exchanged with the single mode,
so the single mode is required to be only the mediator between the two two-level atoms effective interaction.
This effect can not be generated by classical field because classical fields can not couple the two atoms, at any intensity.
When they are initially in a type of Werner state, the entanglement is in general zero at high energy,
but the so called \textit{E-beats} effect take place and the narrow beats are separated by the EDVs \cite{Eberly}.
In other words, in that regime, the initial entanglement amount is periodically recoverd in a sudden manner, only for a short moments separated by the time scale $\pi\Delta/g^2$.
The wide of a E-beat is inversely proportional to the energy of the single mode.
Besides, in that atomic initial condition the entanglement does not change when the single mode is initially in the vacuum state.
However, we have already seen that the vacuum initial state makes an important effect when each atoms starts in a pure state.

A physical implementation of this Hamiltonian interaction between two two-level system and a single mode can be performed
with two quantum dots interacting with a boson mode \cite{Krummheuer}.
Another physical implementation could be implemented considing the Zeeman's level structure in a
$^{138}$Ba cold ion moving in a linear Paul trap \cite{Raizen} in a standing wave configuration \cite{Cirac}.
Spontaneous emission is suppressed using as a qubit the $S_{1/2}$
ground and the $D_{5/2}$ upper metastable states \cite{Blatt,Kli}.
The lifetime of those metastable states of Ba$^+$ is about $45 s$.
The motion of the ions can be described in terms of the normal center-of-mass mode.
The required dispersive interaction between two ions with the same center-of-mass mode can be always simulated.

\begin{acknowledgments}
The authors thank to M. L. Ladr\'{o}n de Guevara, P. Toschek, and P. Zoller for valuable comments. R. P.-G. thanks to G. A. Olivares-Renter\'{\i}a.
This work was supported by Grants: Milenio ICM P02-49F and FONDECyT No. 1030671.

\end{acknowledgments}

\end{document}